\RequirePackage{fix-cm}
\documentclass[smallextended]{svjour3}       
\smartqed  
\usepackage{graphicx}
\journalname{EPJ Sciences 4Open}

\begin{document}

\title{The Origin of the Elements and Other Implications of Gravitational Wave Detection for Nuclear Physics
}

\titlerunning{Gravitational Waves and Nuclear Physics}        
\author{D. Lunney\\~\\
\it{The wave that launched a thousand shifts}}  
\authorrunning{D.~Lunney} 
\institute{ D. Lunney \at
Institut National de Physique Nucl\'eaire et de Physique de Particules du CNRS, \\
Universit\' e Paris-Saclay \\
~\\
            corresponding author \email{david.lunney@ijclab.in2p3.fr}  
}

\date{Received: \today / Accepted: date}

\maketitle

\begin{abstract}

The neutron-star collision revealed by the event GW170817 gave us a first glimpse of a possible birthplace of most of our heavy elements. The multi-messenger nature of this historical event combined gravitational waves, a gamma-ray burst and optical astronomy of a ``kilonova,'' bringing the first observations of rapid neutron capture ($r$ process) nucleosynthesis after 60 years of speculation. 
Modeling the $r$ process requires a prodigious amount of nuclear-physics ingredients: practically all the quantum state and interaction properties of virtually all neutron-rich nuclides, many of which may never be produced in the laboratory! 
Another essential contribution of nuclear physics to neutron stars (and their eventual coalescence) is the equation of state (EoS) that defines their structure and composition. The EoS, combined with the knowledge of nuclear binding energies, determines the elemental profile of the outer crust of a neutron star and the relationship between its radius and mass. In addition, the EoS determines the form of the gravitational wave signal.  This article combines a tutorial presentation and bibliography with recent results that link nuclear mass spectrometry to gravitational waves via neutron stars.

\end{abstract}

\section{Introduction - the elements and their origin}
\label{intro}

The year 2019 was designated the ``International Year of the Periodic Table of Chemical Elements (IYPT2019)'' by the United Nations General Assembly and UNESCO \cite{iypt} as it marked the 150th anniversary of the formulation of the periodic table, by Dmitri Mendeleev.  By organizing the known elements into rows and columns (``periods'') Mendeleev was able to predict new elements, based on available spaces in his table.  Though he gave no explanation as to their origin,  Mendeleev bridged the gap between alchemy and modern chemistry with his famous elementary work \cite{dream}.  

The simplest of all elements, hydrogen (a single proton) was formed after the Big Bang.  The first minutes of the early Universe saw the era of Big-Bang Nucleosynthesis (BBN), which produced the next element:  helium (two protons that have ended up with one or two neutrons) and a little bit of lithium (three protons, four neutrons).\footnote{For BBN, see the historical reference \cite{Wag}, the review by Schramm and Turner \cite{ST} and a recent review  by Coc and Vangioni \cite{CV} that nicely articulates the associated nuclear physics.}   
We now know that rest of the elements are forged by nuclear reactions in stars, where hydrogen fuses over billions of years to produce helium.  Helium nuclei ($^{4}$He) then combine to form $^{8}$Be, which is essentially unstable (its half-life is less than $10^{-16}$ seconds).  The astrophysicist Fred Hoyle reasoned that to overcome this pitfall, the carbon nucleus had to have a quantum state that could accommodate the rare encounter of three helium nuclei.  This brilliant deduction was verified experimentally by William Fowler, who later received the Nobel prize for his work.\footnote{Having predicted the state in carbon, it is tempting to wonder why Hoyle did not share the Nobel prize.  This is a story of much speculation.
Though the late Hoyle would not have been comforted by the fact, the state in question is named after him - the only nuclear state bearing someone's name!
}

Fowler and Hoyle teamed up the astronomer duo Margaret and Geoffrey Burbidge to produce the seminal work (known as B2FH, after the initials of the authors' last names) on the orgin of the chemical elements in stars \cite{B2FH}.\footnote{Remarkably, another treatise on ``nucleogenesis in stars'' was written the same year, by the Canadian physicist Alistair Cameron \cite{Cameron} but is seldom cited since it originally appeared as a Chalk River  Laboratory report.}   
B2FH exhaustively outlined the origins of different groups of isotopes, notably $s$- and $r$-process isotopes whose abundances reflect the same reaction (radiative neutron capture) but different rates (i.e. $slow$ and $rapid$).  When a nucleus is too neutron rich it beta decays (by which a neutron is converted to a proton) back towards stability, thus moving the flow of mass to heavier elements.  Supernova were thought to be the astrophysical site of the $r$ process, given the high temperature, high neutron density and the explosion that copiously enriches the interstellar medium.  But modeling supernova has proven extremely difficult with only recent success in making them explode \cite{explode}.  A very recent review of the $r$ process (and ``astronuclear'' physics in general) by Arnould and Goriely \cite{arnould} includes descriptions of the various possible sites, among them neutron-star collisions -- the landmark event concerning this article.  This site was proposed by Lattimer and Schramm \cite{LattimerSchramm} and later elaborated in \cite{Li,Ross,Frei}. 

\subsection{Nuclear Binding Energy}

As outline above, it is nuclear reactions that power the stars and forge the elements.  It is therefore not surprising that modeling the various nucleosynthesis processes requires an enormous amount of nuclear-physics input data.  The $r$ process in particular requires knowledge of the ground-state properties of practically all neutron-rich nuclei that can exist!  

Mass measurements are particularly important for the study of nuclear structure since they provide the nuclear binding energy, as a consequence of the famous relation $E=mc^2$.  The pioneering precision mass spectrometry of Francis Aston \cite{aston} during the 1920s established that the mass of the atom was about 1\% smaller than the sum of the masses of its constituents, the difference reflecting the binding energy.  Aston also went on to resolve the so-called whole-number rule that lead to the classification of isotopes and received the Nobel Prize for Chemistry in 1922.  
Aston's more famous contemporary, Arthur Eddington, realized that the nuclear binding energy could account for the energy output of stars, which could not be reconciled with their ages considering chemical combustion \cite{edd}.  The binding energy determines the amount of energy available for a given reaction involved in nucleosynthesis i.e. the neutron captures and beta decays that constitute the $r$ process.  Thus the link between mass spectrometry and the fusion of the elements in stars dates from the beginning of the field.  Aston went on to measure the masses of over 200 stable isotopes, which resulted in his so-called ``packing fraction'' \cite{aston}.  This binding energy per nucleon is relatively constant and gave a first clue for the short-range nature of the nuclear interaction.  The peak near the region of $^{56}$Fe strongly correlates with the abundance peak of the well-bound iron-group elements.  

\begin{figure*}
\includegraphics[width=1.0\textwidth]{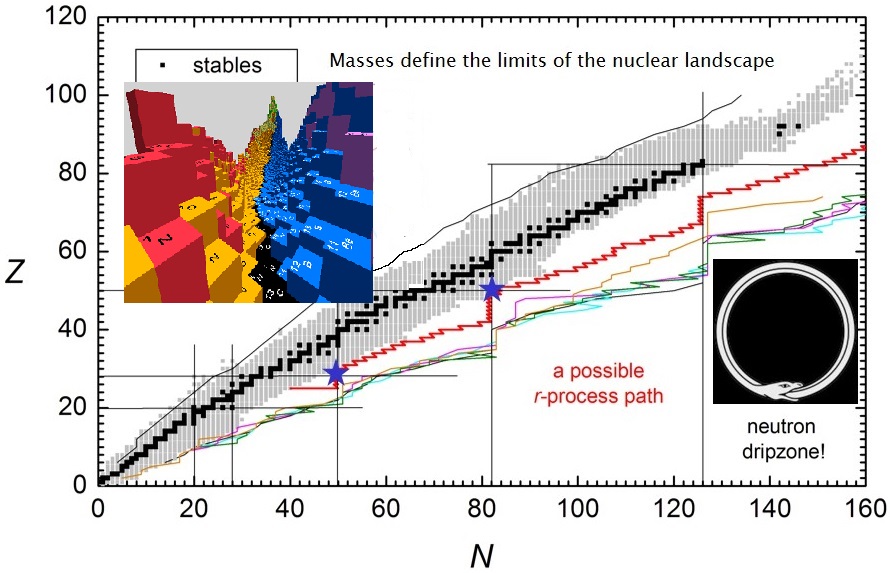}
\caption{Chart of nuclides, formed by plotting atomic number $Z$ as a function of neutron number $N$ for known radioactive isotopes (grey).  Stable isotopes are shown by black squares. Horizontal and vertical lines indicate magic $Z$ and $N$, respectively (see text) and the stars indicate the doubly magic nuclides $^{78}$Ni and $^{132}$Sn.  The red line shows a possible $r$-process path and the colored lines correspond to neutron drip lines calculated by different mass models.  Inset (left) is a surface plot defined by the mass excess, derived from measured binding energies.  Beta-decaying nuclides are in orange and blue, alpha-emitters in yellow, proton-emitters in red and neutron emitters in purple.  Stable nuclides, again in black, are seen to form the so-called valley of stability into which all nuclides with mass excess decay.}
\label{surface}       
\end{figure*}

The advent of the particle accelerator in the 1930s allowed the synthesis of radioactive isotopes, masses of which could be linked to the measured reaction energies.  Today, masses of over 3000 isotopes have been measured.  These are indicated in Fig.~\ref{surface}, called the chart of nuclides.  

The horizontal and vertical lines in Fig.~\ref{surface} illustrate the so-called ``magic'' numbers, which correspond to filled nuclear shells, in analogy with the atomic shell model.  Shell closures were  identified by discontinuities in mass differences by Elsasser \cite{els} and amongst the most important questions in nuclear physics is whether these shell closures lose their stabilizing power for the exotic doubly magic systems  $^{78}$Ni and $^{132}$Sn.  Details of nuclear shell structure from the mass surface are discussed at length in a 2003 review paper \cite{rmp}. 


Masses can be measured by a variety of techniques that are often adapted to the production mechanisms of the nuclides in question.  Details of these techniques and production facilities are also discussed in the 2003 review \cite{rmp}.  For the last 30 years, the technique of Penning-trap mass spectrometry has taken over the field due to its inherent accuracy.  The original review \cite{rmp} introduces Penning traps and the facilities in operation today.  Updates \cite{update} and more details can be found in the 2013 Special Issue on the 100 Years of Mass Spectrometry \cite{100}.  

As seen in Fig.~\ref{surface}, a possible $r$ process path runs through regions of the chart where nuclides have yet to be produced in the laboratory.  In the neutron-rich environment of a neutron-star merger, the $r$ process would even run along the limit of nuclear stability:  the neutron drip line.  Studies of nucleosynthesis have therefore no recourse but to turn to nuclear theory.  

\subsection{Nuclear Mass Models}
\label{models}

One of the major challenges to nuclear theory is to consistently model nuclear sizes, shapes and excited states while predicting the associated binding energy.  With good mass values comes a reliable prediction of the limits of nuclear stability, which are defined by the nucleon-separation energies, derived from the mass (see \cite{rmp}). 
Attempts at predicting mass values analytically date from the beginning of nuclear physics, the most well known being the Bethe-Weiszaecker formula (see \cite{rmp} for a comparison of the different types of formulas and models).  Such formulas contain many parameters that are adjusted to measured masses so that near extrapolations can be made.  Between 6000 and 8000 nuclides are predicted to exist as bound systems, depending on the model used.  However such phenomenological formulas could not provide all of the quantities necessary for modeling the $r$ process so a mass model based on a nuclear interaction was needed.

Beginning with HFB-1 in 2002 \cite{hfb1}, the Brussels-Montreal group has constructed a series of global mass models based on the Hartree-Fock-Bogoliubov method using Skyrme-type nuclear energy-density functionals. 
This landmark achievement included a systematic study of the different force parameters (themselves adjusted to other nuclear properties) elaborated through almost 20 publications, reaching HFB-32 \cite{hfb32}.  This approach was also extended to the Gogny force, D1M \cite{D1M} and also to the relativistic mean field \cite{RMF}.  The HFB models now calculate masses with accuracies that rival phenomenological formulas, but additionally provide the self-consistent framework for calculating all the nuclear properties required by $r$-process modeling.  

Along the way, the Brussels-Montreal collaboration introduced a constraint of particular interest for predicting the properties of neutron-rich nuclides:  the force calculating the masses also reproduces an equation of state (EoS) describing that of neutron matter \cite{hfb9}.\footnote{The neutron-matter constraint was first introduced in 1972 by Ravenhall, Bennett and Pethick \cite{RBP} to improve an existing Skyrme force and later used by Chabanat et al. \cite{Sly230} for elaborating the ``Lyon'' Skyrme force but the first time that an interaction was co-constrained by nuclear masses and neutron matter was in 1982 by the Brussels forefathers Rayet, Arnould, Tondeur and Paulus (the force was called RATP) \cite{RATP}.}  The authors of HFB-9 \cite{hfb9} made it clear that the interest of this constraint was for modeling an $r$ process that might occur during the decompression of neutronic matter after a neutron-star merger.  Another interesting addition to the suite of HFB models was the self-consistent inclusion of fission barriers with HFB-8 \cite{hfb8} since it now accepted that fission recycling plays an important role in shaping the $r$-process abundances - see \cite{spy} for the state of the art.  The EoS and its crucial impact for neutron stars and gravitational waves is discussed below.   


\section{Neutron Stars}

Born from the cataclysmic supernova explosions of massive stars, neutron stars are very probably the densest objects in the Universe.  The super-nuclear density is such that the mass of the sun would be squeezed into a sphere with the diameter the size of a city.

Neutron stars were hypothesized by the formidable duo of Walter Baade and Fritz Zwicky in 1932, barely a year after the discovery of the neutron.\footnote{In a very interesting science-history paper, Yakovlev et al. \cite{Landau} describe the visionary work of Lev Landau, whom they believe imagined compact stars with nuclear densities even before the neutron's discovery.}  But the work of these astronomers is impressive for another reason:  in their 1933 communication \cite{aps}, Baade and Zwicky introduced a new class of astronomical objects they called supernovae.  After describing this major discovery (which they also linked to cosmic rays), they made a further prediction:  ``With all reserve we advance the view that supernovae represent the transitions from ordinary stars  into {\it neutron stars}...'' (emphasis theirs).  Perhaps it is natural that the inventors of supernovae and neutron stars also linked them.  In 1939, Robert Oppenheimer produced the landmark paper \cite{OV} containing the first detailed calculations of neutron-star structure using the powerful concepts brought by General Relativity.  The same year Oppenheimer continued his work on gravitational contraction, publishing the first detailed description of black-hole formation \cite{OS}.\footnote{These wonderful discoveries are presented along side the mathematical detail they merit in the venerable book called {\it Gravitation} by Misner, Thorne and Wheeler \cite{gravitation}.  They also mention an astounding pronouncement made by Laplace in his 1795 {\it Exposition du Syst\`eme du Monde} \cite{laplace} that stars with enough mass would have so much gravitational pull that light could not escape.  In fact, the unsung English natural philosopher and clergyman John Michell proposed the same idea, referring to ``dark stars'' in a 1784 paper for the Royal Society of London \cite{michell}. Their reasoning predates the discovery of black holes by LIGO/Virgo by over two centuries!}

The initial rotation of a large star gives rise to enormous angular momentum in the resulting compact object.  Though composed primarily of neutrons, electrons and protons also present give rise to a strong magnetic field that traps charged particles, causing radio emission.  Since the magnetic poles can differ from the rotation axis, neutron stars blink like radio beacons.  The pioneering radioastronomy experiments of Hewish and Bell-Burnell detected these signals as so-called pulsars in 1967 \cite{HB}.   

\subsection{Neutron-star composition}
\label{nsc}

Modeling neutron stars requires a broad range of physics, given the extreme environment.  
These compact objects are not burning and in spite of their violent birth, cool very rapidly via neutrino emission.  Cold neutronic matter is thus assumed to be in beta equilibrium and with increasing pressure, electron capture and neutrino emission make the composition more neutron rich.  A concise and authoritative work on neutron-star structure is that of Lattimer and Prakash \cite{LP} with more recent reviews by Vida\~na \cite{vidana} and by Blaschke and Chamel \cite{BC}.  

Most of our knowledge of neutron stars comes from mass measurements of binary systems, involving pulsars (see \cite{ozel}).  Oppenheimer and Volkoff \cite{OV} calculated a neutron-star mass of 0.72 solar masses using a free-neutron gas model.  The much heavier (1.4-solar-mass) binary pulsar discovered in 1974 by Hulse and Taylor \cite{HT} was the first proof that nuclear forces must play a role by making the EoS more rigid.\footnote{It is interesting to note that monitoring the Hulse-Taylor pulsar subsequently revealed a slow decrease in rotation frequency that corresponded to the emission of gravitational waves - indirect evidence, but another victory for Einstein's General Relativity.}  From the many observations \cite{ozel} the canonical neutron star is about 1.4 solar masses.  The recent detections of two-solar-mass neutron stars has placed an even stronger constraint on the nuclear EoS.  We return to this important question in section \ref{eosnm}.  

The outer layers of a neutron star are thought to consist of a solid Coulomb lattice of neutron-rich nuclei, forming a crust.  The core may contain hyperonic matter or even deconfined quarks, however this is territory rich in speculation.  Between the crust and the core, competition between Coulomb repulsion and nuclear attraction in the cold crystalline lattice can cause the formation of interesting geometries that have been described as nuclear ``pasta'' (see the recent review \cite{horowitz} that offers an enticing menu of gnocchi, spaghetti, lasagna and other exotic forms).   

The outer crust forms  below the atmosphere, essentially from the surface, reaching a depth close to nuclear-matter density:  about $2\times 10^{8}$~g/cm$^{3}$.  Deeper in the crust, increasing pressure brings more neutrons into play until the drip-line density ($4\times 10^{11}$ g/cm$^{3}$, or $2.5\times 10^{-4}$ nucleons/fm$^{3}$) is reached, marking the transition from the outer to inner crust.  
The increasing density causes a stratification of the outer crust with deeper layers containing more neutron-rich nuclides.  The composition is determined by minimizing the Gibbs free energy per nucleon at a given pressure.   
All other input (lattice energy per cell, mean electron energy density, pressure and number density) is relatively robust and well known so that the composition depends essentially on the nuclear binding energy.  Also required is the EoS, which describes the variation of pressure as a function of density (for cold nuclear matter).  To see how the composition, pressure and density vary with depth, the TOV equations \cite{tov,OV} are integrated using the appropriate EoS from the surface towards the center.  
The physics and associated modeling of the neutron-star crust and its composition are explained in detail by Chamel and Haensel \cite{CH}.    

Observations have revealed the presence of $^{62}$Ni and $^{56}$Fe in the (very thin) atmospheres of neutron stars.  As one ``drills'' deeper into the neutron-star the increasing pressure favors the presence of heavier nuclides.  The original theoretical drilling explorations of Tondeur \cite{Tondeur} and more often-cited\footnote{possibly since Tondeur's work was published in French.} Baym et al. \cite{bps} were extended by Pearson, Goriely and Chamel \cite{Pearson2011,Goriely2011}, using the Brussels Skyrme forces with their self-consistent masses and EoS described earlier.  Ruester et al. \cite{Ruester2006}, Roca-Maza and Piekarewicz \cite{Roca}, and Kreim et al. \cite{Kreim2013} have also pursued such work, studying the impact of a wider range of mass models, while Utama et al. \cite{Utama} have done so using a Bayesian neural network approach.  Chamel \cite{Chamel2020} has now greatly improved the speed of such calculations by avoiding a full minimization of the Gibbs free energy, which allows more systematic study and better determination of the abundances in the thinner but deeper layers of the star.  Also discussed in \cite{Chamel2020} is the treatment of the inner crust, which is important for modeling the ejection of $r$ process nuclides from neutron-star mergers.  

\begin{figure*}
\includegraphics[width=0.65\textwidth]{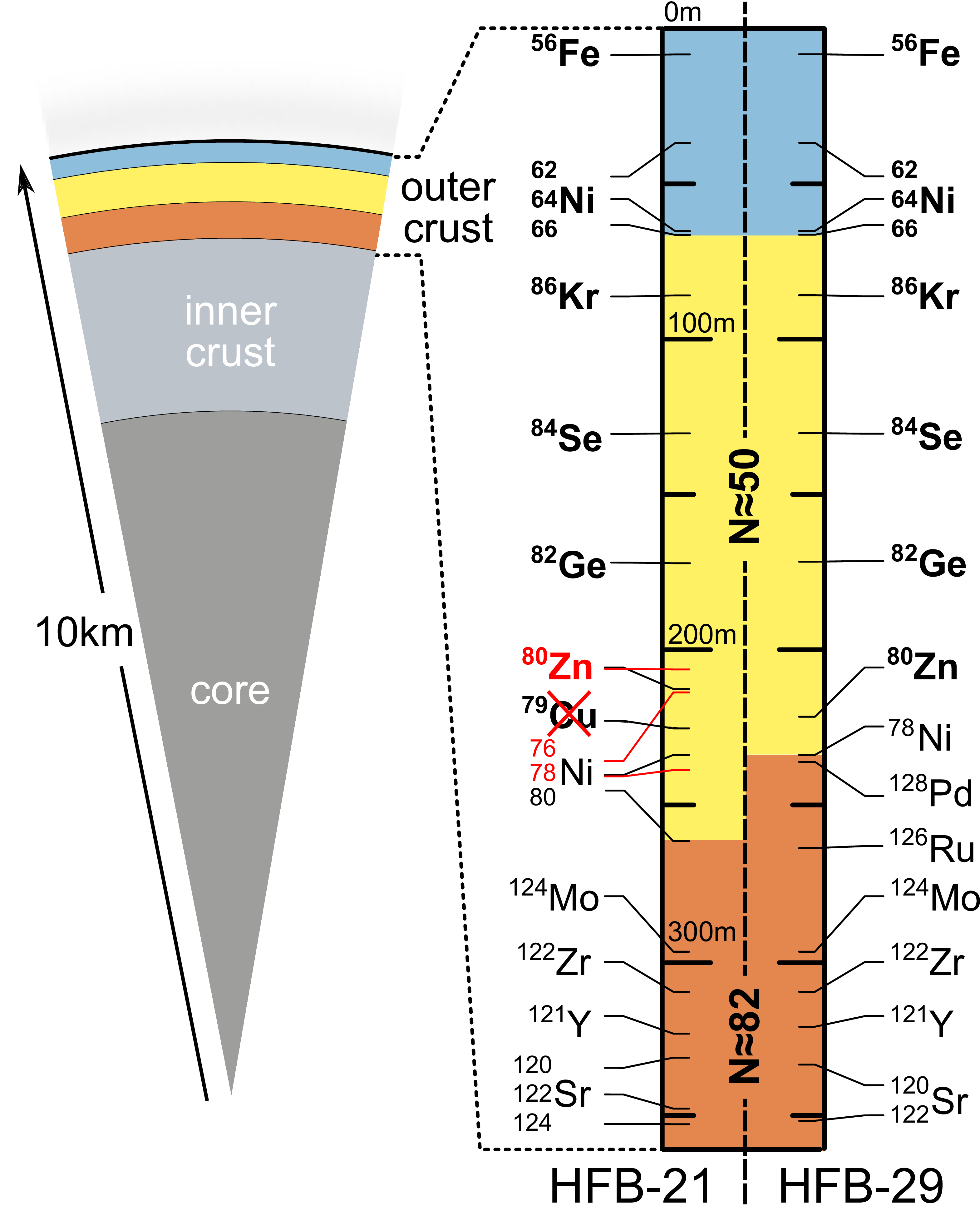}
\caption{Depth profile of a 1.4-solar-mass, 10-km radius neutron star showing predictions of microscopic mass models HFB-21 and HFB-29 for the outer crust calculation described in the text.  The binding energies of the nuclides in bold face have been determined experimentally.}
\label{ns} 
\end{figure*}

The Brussels-Montreal studies with HFB-19, 20 and 21 \cite{hfb192021} predicted that deeper layers of neutron stars would bear nuclides near the $N=50$ neutron shell closure, such as $^{78}$Ni and $^{82}$Zn \cite{Pearson2011}.  The prediction of $^{82}$Zn by BSk19 was accompanied by an ISOLTRAP mass measurement of this exotic nuclide by Wolf et al. \cite{Wolf}.  The measured value turned out to be more bound than the HFB-19 prediction, so that updated neutron-star-composition calculations saw it removed it from the crust, replacing it with $^{80}$Zn.  Wolf et al. \cite{Wolf} also presented a profile calculated using HFB-21, which predicted the presence of $^{79}$Cu, an odd-$Z$ nuclide which would normally be less bound.  ISOLTRAP later measured the mass of the exotic $^{79}$Cu \cite{Welker}, finding it more bound than the HFB-21 prediction so that it, too disappeared from the crust.  Fig.~\ref{ns} shows the neutron-star profile from \cite{Wolf} updated with the $^{79}$Cu mass \cite{Welker} with HFB-21 and the newer HFB-29 \cite{hfb32} models for unknown masses.  A more recent study of the symmetry energy using the mass models based on BSk22 and BSk 24-26 was performed by Pearson et al. \cite{Pearson2018}, showing crustal compositions and including a discussion of the case of $^{79}$Cu (and odd-$A$ nuclides in general).  For the case of Fig.~\ref{ns}, the outer crust composition is now experimentally constrained to about 240 m and the predictions deeper in the crust are quite consistent.  Mass measurements therefore still play a crucial role for constraining composition and the development of models.    

\subsection{The Equation of State (EoS) of neutronic matter}
\label{eosnm}

\begin{figure*}
\includegraphics[width=0.99\textwidth]{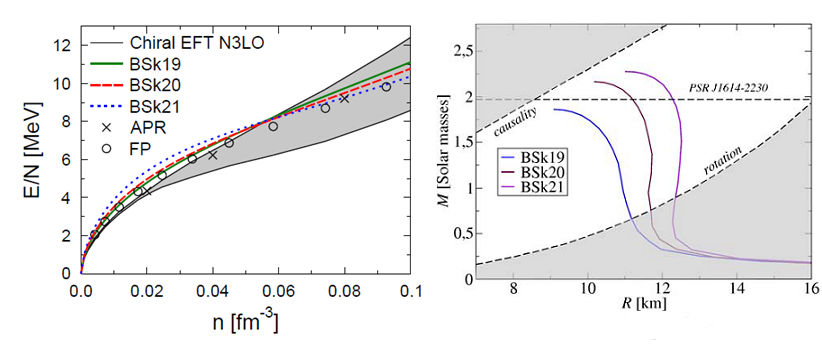}
\caption{(left) The Equation of State (EoS) of neutron matter: energy density $E/N$ as a function of neutron density $n$ showing different extrapolations towards neutron-star density. (right) Neutron-star mass versus radius plots solved using the TOV equations with the three different EoS from BSk-19, 20, and 21 (left) as input \cite{Chamel2011}.  The discovery of the two-solar-mass neutron star in the binary pulsar J1614-2230 \cite{Demorest,Fonseca} rules out the EoS of BSk19, as do J0348+0432 \cite{Antoniadis} and J0740+6620 \cite{Cromartie}.}
\label{eos} 
\end{figure*}

We described in section \ref{models} the interesting EoS constraint imposed on the Brussels Skyrme forces.
With HFB-9 \cite{hfb9} this was done using the Friedman-Pandharipande (FP) calculation, the form of which is shown in Fig.~\ref{eos} (left).  Other EoS are also shown in the figure, including those corresponding to the formulations of the Brussels Skyrme forces BSk19, BSk20, and BSk21 \cite{hfb192021}.  The forces were in fact constructed with increasing stiffness for a systematic EoS study, reported by Chamel et al. \cite{Chamel2011}.  

We have mentioned pressure within the neutron-star crust and the resulting composition related to depth using the Tolman-Oppenheimer-Volkoff (TOV) equations, derived from the theory of general relativity \cite{tov,OV}.  Integrating the TOV equations using a given EoS for a neutron star (here considered spherical and non rotating) provides an important neutron-star characteristic called the gravitational mass-radius plot.  Such a plot is shown in the right side of Fig.~\ref{eos}, with the corresponding EoS shown in the left panel.  The definition of the limiting areas of rotation and causality is nicely described by Lattimer \cite{Lattimer2012} who also illustrates how the inflection of the EoS corresponds to the mass-radius behavior.  Lattimer has also updated this work in light of GW170817 \cite{Lattimer2020}.  The reader is also referred to the website by M.~Hempel \cite{Hempel} that has the data available for a large family of EoS as well as the corresponding mass-radius plots.   

The right panel of Fig.~\ref{eos} shows a horizontal dashed line that corresponds to the 1.97-solar-mass neutron star (J1614-2230) reported by Demorest et al. \cite{Demorest} (this mass value was updated to 1.928 by Fonseca et al. \cite{Fonseca}). Further neutron-star mass measurements have reported 2.1 solar masses for J0348+0432 \cite{Antoniadis} and 2.14 solar masses for J0740+6620 \cite{Cromartie}.  We see that the softness of the BSk19 force describes neutron stars that are too light.  Thus, the interest of laboratory nuclear physics on constraining neutron-star masses is nicely illustrated.\footnote{A very recent detection by the LIGO-Virgo Collaboration (GW190814 \cite{gw190814}) reveals a compact object of 2.6 solar masses, which would be either the lightest black hole or the heaviest neutron star ever discovered!  The latter would exclude all the EoS shown in Fig.~\ref{eos}.}
  
The more recent work of Pearson et al. \cite{Pearson2018} mentioned above in the context of the crustal composition also examines the EoS correspond to BSk22 and BSk24-26 and shows the corresponding neutron-star mass-radius plots with ample discussion.  Moreover, the neutron-star descriptions based on BSk20-26 are all compatible with the mass-radius constraints obtained from the observations associated with GW170817, as illustrated in \cite{PCS}.  This includes the extremely important tidal deformability coefficient that depends strongly on the EoS.  
A critical examination of these effects was made by Tews, Margueron and Reddy \cite{Tews}, who concluded that the information brought by GW170817 does not yet constrain our current knowledge of the EoS from nuclear physics.  Tews, Margueron and Reddy also address the interesting possibility of phase transitions occurring in neutron-star cores (not considered in the Brussels EoS).  Improvements in gravitational wave detection are expected to bring such knowledge in the future.

Characterizing gravitational wave events are the (early) inspiral phase and the (later) ringdown stage.  Neutron-star mergers have particularly long inspiral times compared to black-hole mergers.    Fitting these detected signals critically depends on the choice of nuclear equation of state.  
A recent study illustrating this and how it can be used in future detections is given by Bauswein et al. \cite{Bauswein}.  
A more general source for the EoS and its role in describing compact objects is the recent review article by Oertel et al. \cite{Oertel}.  

\section{The multi-messenger events associated with GW170817}

The extraordinary neutron-star merger event GW170817 (it occurred August 17, 2017) was announced with great excitement \cite{press}.  After the initial gravitational wave detection by LIGO, additional localization information from the European interferometer Virgo enabled the world's astronomers to train their space- and Earth-based telescopes towards the stellar drama that was unfolding in the Hydra constellation of the southern sky, 130 million light-years away.  

This story is documented by a remarkable publication, heralding the birth of multi-messenger astronomy \cite{gw170817}.  The article assembles the largest-ever astronomy author list, from some 70 different observatories (seven satellite born) covering the entire electromagnetic spectrum and every continent, even Antarctica with the IceCube neutrino detector (which did not see any events).  Observations lasted for weeks afterwards.  The reader is also referred to a Focus-Issue collection of publications on the electromagnetic counterpart of GW170817 \cite{focus}.

\subsection{The gamma-ray burst and the ``kilonova''}

The electromagnetic observations triggered by GW170817 reached from radiofrequency (at later observation time) all the way past x-rays to gamma rays.  
Notification of a gamma-ray burst GRB170817A was in fact given six minutes earlier than GW170817, by the Fermi satellite \cite{gw170817}.\footnote{When the exact timings were compared, the speed of gravitational-wave propagation was found to be consistent with the prediction of General Relativity - another victory for Einstein!}  GRB170817A was also detected by the SPI instrument of the INTEGRAL satellite through an off-line search initiated by LIGO-Virgo and Fermi, helping further constrain the localization.

As mentioned, GW170817 and GRB170817A triggered an unprecedented observation campaign.  Though the location was well defined, it was not visible to most of the large optical telescopes until 10 hours after the merger.  The one-meter Swope telescope at Las Campanas, Chile was the first to observe (and announce) the bright optical transient AT2017gfo.  The telescope is used in the Swope Supernova Survey program and could not rule out a faint supernova.  It took further observations at different wavelengths to confirm an event only seen once before:  the kilonova.  

The kilonova is a transient whose energy release is inferior to that of a supernova but superior to that of a nova explosion (novae occur during the burning of hydrogen accreted by a white dwarf star in a binary system).  Kilonovae were hypothesized to appear from the radioactive-decay heat created by $r$-process ejecta \cite{Li,Rosswog2005,Metzger2010}.  The first kilonova was discovered by Tanvir et al. \cite{Tanvir} and Berger et al. \cite{Berger} using the Hubble Space Telescope, also triggered by a $\gamma$-ray burst.  This was a tantalizing event since it evoked two exciting possibilities:  that compact-object mergers are the progenitors of short-duration $\gamma$-ray bursts and also the sites of significant production of $r$-process elements.  Tanvir et al. \cite{Tanvir} went on to suggest that kilonovae could offer an alternative electromagnetic signature for direct detection of gravitational waves.  This bold statement was vindicated by GW170817.  

A recent review of kilonovae is given by Metzger \cite{Metzger2017}, who concludes that the largest uncertainties are related to the wavelength-dependent opacity of the ejecta, particularly for lanthanide and actinide isotopes as their spectra and ionization states not measured and impossible to calculate reliably.  The optical emission of the AT217gfo kilonova is discussed by Arcavi et al. \cite{Arcavi2017} and the modeling of the ejecta is reported by Kasen et al. \cite{Kasen2017}, who infer the presence of distinct isotopic components heavier and lighter than $A=140$, in sufficient quantities to account for the dominant contribution of $r$-process elements.  Recently, Watson et al. \cite{Watson2019} have argued that lines of strontium ($Z=38$) are discernible in the AT217gfo spectra.  A comprehensive account of neutron-star merger nucleosynthesis performed under state-of-the art hydrodynamical conditions is provided by Just et al. \cite{Just}.

\section{Summary}

The extraordinary event GW071707 has established a long-sought link between neutron stars and the $r$ process. As such, the role of low-energy nuclear physics in the field a gravitational waves can be highlighted.  This has been done using examples of mass measurements with the Penning-trap spectrometer at CERN's ISOLDE facility, combined with new theoretical approaches from the Brussels-Montreal collaboration based on empirical nucleon-nucleon interactions that link to the nuclear equation of state, which is of critical importance for describing neutron stars and the gravitational wave signal of GW170817 detected by LIGO/Virgo.  

~\\
\textbf{Conflict of Interest:}  I have no financial conflict of interest in connection with this article.

\begin{acknowledgements}
Warm thanks to N. Chamel, S. Goriely and J.M. Pearson for edifying discussions and the calculations shown in Figs. 2 and 3,
to R.N. Wolf for producing Fig. 3 
and to the ISOLTRAP Collaboration for the masses of $^{79}$Cu and $^{82}$Zn (among many other measurements).  
\end{acknowledgements}


\end{document}